\documentclass[twocolumn, aps, prd, superscriptaddress]{revtex4-1}  

\usepackage{amsmath,amssymb}
\usepackage{graphicx}
\usepackage[english]{babel}
\usepackage{float}

\begin{document}

\title{Modelling Thermoelastic Distortion of Optics Using Elastodynamic Reciprocity}

\author{Eleanor King}\email{Corresponding author: eleanor.king@adelaide.edu.au}
\affiliation{Department of Physics, The University of Adelaide, Adelaide SA 5005, Australia}

\author{Yuri Levin}
\affiliation{School of Physics, Monash University, Clayton VIC 2800, Australia}

\author{David Ottaway}
\affiliation{Department of Physics, The University of Adelaide, Adelaide SA 5005, Australia}

\author{Peter Veitch}
\affiliation{Department of Physics, The University of Adelaide, Adelaide SA 5005, Australia}

\begin{abstract}Thermoelastic distortion resulting from optical absorption by transmissive and reflective optics can cause unacceptable changes in optical systems that employ high power beams. In “advanced”-generation laser-interferometric gravitational wave detectors, for example, optical absorption is expected to result in wavefront distortions that would compromise the sensitivity of the detector; thus necessitating the use of adaptive thermal compensation. Unfortunately, these systems have long thermal time constants and so predictive feed-forward control systems could be required - but the finite-element analysis is computationally expensive. We describe here the use of the Betti-Maxwell elastodynamic reciprocity theorem to calculate the response of linear elastic bodies (optics) to heating that has arbitrary spatial distribution. We demonstrate using a simple example, that it can yield accurate results in computational times that are significantly less than those required for finite-element analyses. 
	
\end{abstract}


\maketitle 

\section{Introduction}

Advanced-generation interferometric gravitational wave detectors, such as Advanced LIGO \cite{2015advancedligo}, Advanced Virgo \cite{virgo2009advanced} and KAGRA \cite{somiya2012detector} are currently being commissioned. Their sensitivity is expected to surpass that achieved by first generation instruments by almost an order of magnitude in the high frequency region. To achieve this, very high circulating power levels (0.5-1 MW) will be stored within the Fabry-Perot arm cavities. At these power levels, even low levels of optical absorption can lead to significant thermoelastic distortion of optical surfaces and unacceptable levels of wavefront distortion \cite{lawrence2003active}, resulting in reduced circulating power and a reduction in the efficiency of the detector signal readout. Thermally actuated compensation systems will be thus used to ameliorate the wavefront distortion. However, the thermal time constants for the absorption-induced distortion and the compensation are long, typically 12 hours, and thus incorporating predictive modeling in the control systems may prove essential.

The response of a linear elastic system to heating is described by the theory of thermo-elasticity and its applications to highly symmetric, idealized systems are described in many books (see \cite{boley2012theory} for example). It has also been used to develop analytic expressions for less idealized optical systems \cite{hello1990analytical,lawrence2003active}. The expressions developed by Hello and Vinet \cite{hello1990analytical} are relevant to the work described here, but apply only to cylindrical isotropic mirrors heated by coaxial laser beams.

More complicated systems, which incorporate asymmetric heating or anisotropic elasticity, can be investigated using finite-element numerical models that apply the equations of thermo-elasticity on a three-dimensional spatial mesh. For dynamic systems, the thermoelastic equations must be solved at each epoch, requiring computational times that can run to many days. This approach would be untenable for use in predictive feed-forward actuation to control systems. In such cases, the solution of the scalar problem to determine the temperature profile throughout the optic can be solved rapidly; the time consuming part is solving the tensor-based elasticity problem to convert the thermal profile into an elastic distortion.

The Betti-Maxwell theorem of elastodynamic reciprocity \cite{achenbach2006reciprocity} provides an alternative approach to using finite-element methods (FEM) to solve the tensor part of the thermoelastic distortion. It has previously been used to investigate the excitation of Rayleigh-Lamb elastic waves in a metal plate due to heating produced by a line-focused pulsed laser beam assuming that the heating is confined to the surface of the plate and it has infinite lateral extent \cite{achenbach2005thermoelasticity,achenbach2007application}. In the context of gravitational wave detection, it has been used to compute the interferometer's response to creep events in the fibers that suspend the optics \cite{levin2012creep}. We extend its use to predict thermoelastic distortion of an optic of finite size with asymmetric heating.

We describe here how elastodynamic reciprocity and FEM can be combined to provide accurate predictions of thermoelastic surface distortion more quickly than using FEM alone. In summary, FEM is used to determine the response of the optic to a set of orthonormal tractions, or pressures \textemdash a computationally expensive calculation that is performed once for an optic. Then, using reciprocity, the distortion due to the instantaneous temperature profile in the optic is calculated using a sum of scalar volume integrals that incorporate these responses. The computational cost of this step is much less than that of a full elastostatic FEM evaluation. Additionally, it is amenable to parallelization, which would further reduce the computational time.

The layout of the rest of the paper is as follows:
in Section II we introduce the Betti-Maxwell theorem of elastodynamics and show how it can be used to determine the surface distortion by careful choice of a suitable ‘auxiliary’ elastic system. We demonstrate its application by calculating the distortion of the end face of a cylindrical optic that is heated by a Gaussian heat flux that is (a) coaxial with and (b) laterally displaced from the axis. The approach and model are described in Sections III and IV. Finally, the resulting surface distortions are presented in Section V and compared with the results of elastostatic FEM calculations. Computation times for these two approaches are compared in Section VI 

\section{Elastodynamic Reciprocity and Thermal Distortion}

The Betti-Maxwell reciprocity theorem for elastodynamics \cite{achenbach2006reciprocity,achenbach2007application} specifies the relationship between the displacement \(\vec{u}(\vec{r},t)\) that results from an applied surface traction \(\vec{t}(\vec{r},t)\) and internal body force \(\vec{f}(\vec{r},t)\) for two elastic states of a linear elastic body:

\begin{multline} 
\int_{S}{(t_i^2u_i^1-t_i^1u_i^2)}dS
\\
=\int_V\Big[(f_i^1-\rho\ddot{u}_i^1)u_i^2-(f_i^2-\rho\ddot{u}_i^2)u_i^1 \Big]dV
\label{equation1}
\end{multline}

where \(\rho\) is the density, \(\ddot{u}\) is acceleration, the superscripts \(1\) and \(2\) represent the two states, and the Einstein summation convention is us{ed. If \(t_i(\vec{r},t)=t_i(\vec{r})e^{i\omega t}\) and  \(f_i(\vec{r},t)=f_i(\vec{r})e^{i\omega t}\) then \(u_i(\vec{r},t)=u_i(\vec{r})e^{i\omega t}\), and thus Eq. (\ref{equation1}) becomes

\begin{multline}
\int_S \Big( t^2_i(\vec{r}) u^1_i(\vec{r})-t^1_i(\vec{r}) u^2_i(\vec{r})\Big) dS
 \\
 = \int_V \Big(f^1_i(\vec{r}) u^2_i(\vec{r})-f^2_i(\vec{r}) u^1_i(\vec{r})\Big)dV 
\label{equation2}
\end{multline}

 We shall use this theorem to determine the surface displacement (distortion) due to heating of an optic by, for example, partial absorption of an incident laser beam. For the first state, which we shall refer to as the “thermal state” and label \(T\), we assume that the optic is free and thus \(t_i^T=0\),  and there is a non-zero body force due to the heating. Since we are interested in the distortion of the end face of the optic, we choose the second state, which is often referred to as the “auxiliary state” and we shall label \(A\) , to have a traction \(t_z^A\)  applied to the end face of the optic and assume  \(f_i^A=0\). Thus, Eq. (\ref{equation2}) becomes

\begin{multline}
\int_S  t^A_z(\vec{r}) u^T_z(\vec{r})dS 
\\
= \int_V f^T_i(\vec{r}) u^A_i(\vec{r})dV = \int_S  \sigma^T_{ij}(\vec{r}) \varepsilon^A_{ij}(\vec{r})dV 
\label{equation3}
\end{multline}

where \(\varepsilon_{ij}^A(\vec{r})\) is the internal strain produced by the traction \(t_z^A(\vec{r})\) , and \(\sigma_{ij}^T(\vec{r})\)  is the internal stress associated with the body force: \(f_i^T=-\frac{\partial \sigma_{ij}^T}{\partial x_j}\)  . 

Consider now applying time-harmonic tractions with amplitude $t_z^A(\vec{r}_s)=\chi_n(\vec{r}_s)$ $n = 1,2,...$. It is convenient to choose $\chi_n(\vec{r}_s)$ to be
orthonormal, so that $\int \chi_n(\vec{r}_s)\chi_m(\vec{r}_s)dS=\delta_{nm}$.
 Then, expressing the surface displacement amplitude as:

\begin{equation}
u^T_z(\vec{r}) = \sum_{m} a_m\chi_m (\vec{r})
\label{equation4}
\end{equation}

transforms the left-hand term of Eq. (\ref{equation3}) to

\begin{equation}
\int_S t^A_z(\vec{r}) u^T_z(\vec{r}) dS = \int_S \chi_n (\vec{r}) \sum_{m} a_m\chi_m (\vec{r}) dS = a_n.
\label{equation5}
\end{equation}

Therefore 

\begin{equation}
a_n = \int_V \sigma^T_{ij}(\vec{r}) \varepsilon^A_{ij}(\vec{r})dV 
\label{equation6}
\end{equation}

That is, if the amplitude of the elastic response of the optic, \(\epsilon_{ij}^A(\vec{r})\) , to each of the tractions \(\chi_n(\vec{r})\)  is known then the amplitude of the distortion of the end face of the optic, \(u_z^T(\vec{r})\) , due to any thermal stress distribution can be calculated using Eqs. (\ref{equation4}) and (\ref{equation6}).

We shall use this approach to calculate the surface distortion due to non-uniform heating of a homogeneous isotropic body for which

\begin{equation}
\sigma^T_{ij}(\vec{r}) = \frac{-E\alpha}{1-2\nu} \Delta T(\vec{r})\delta_{ij} 
\label{equation7}
\end{equation}

where   \(E \) is Young's modulus, \(\alpha\) is the coefficient of thermal expansion, \(\nu\) is Poisson's ratio, \(\Delta T=T(\vec{r})-T_0\) , and \(T_0\)  is the ambient temperature. Eq. \ref{equation6} thus becomes

\begin{equation}
a_n = \frac{-E\alpha}{1-2\nu} \int_V(T(\vec{r})-T_0) \textrm{Tr}\big\{ \varepsilon^A (\vec{r})  \big\} dV 
\label{equation8}
\end{equation}

\section{Implementation}

To determine the distortion of the end-face using reciprocity, one must first characterize the response of the elastic system, \(\varepsilon_{ij}^A(\vec{r})\) , to a set of orthonormal basis tractions \(t_z^A(\vec{r},t)=\chi_n(\vec{r}) \exp{[i\omega t]}:n = 1, ....,N\)  using an elastostatic FEM [11]. 

Zernike functions would be a tempting choice given our cylindrical geometry, particularly as they are orthogonal to a uniform traction and thus applying the auxiliary tractions should not apply net forces to the system. However, as shown in Section IV, they are not well suited to describing the surface distortion.

The orthonormal basis tractions we shall use apply a non-zero (instantaneous) force to the optic, leading to ill-conditioning of the FEM at very low frequencies. We thus used a traction frequency of \(\omega = 1\) Hz as the response is independent of frequency for frequencies well below the first resonance - see [12] for example.

In all of our numerical tests, we assume a cylindrical fused silica optic with height \(h = \) 200 mm , radius \(R =\) 170 mm , \(E =731\) MPa, \( \nu = 0.17 \) and \(\alpha = 0.55 \times 10^{-6} \) K\textsuperscript{-1}   . A radial cross section of the optic and the meshing used for the FEM is shown in Fig. \ref{figure1}.

\begin{figure}[htbp]
	\centerline{\includegraphics[width=0.9\columnwidth]{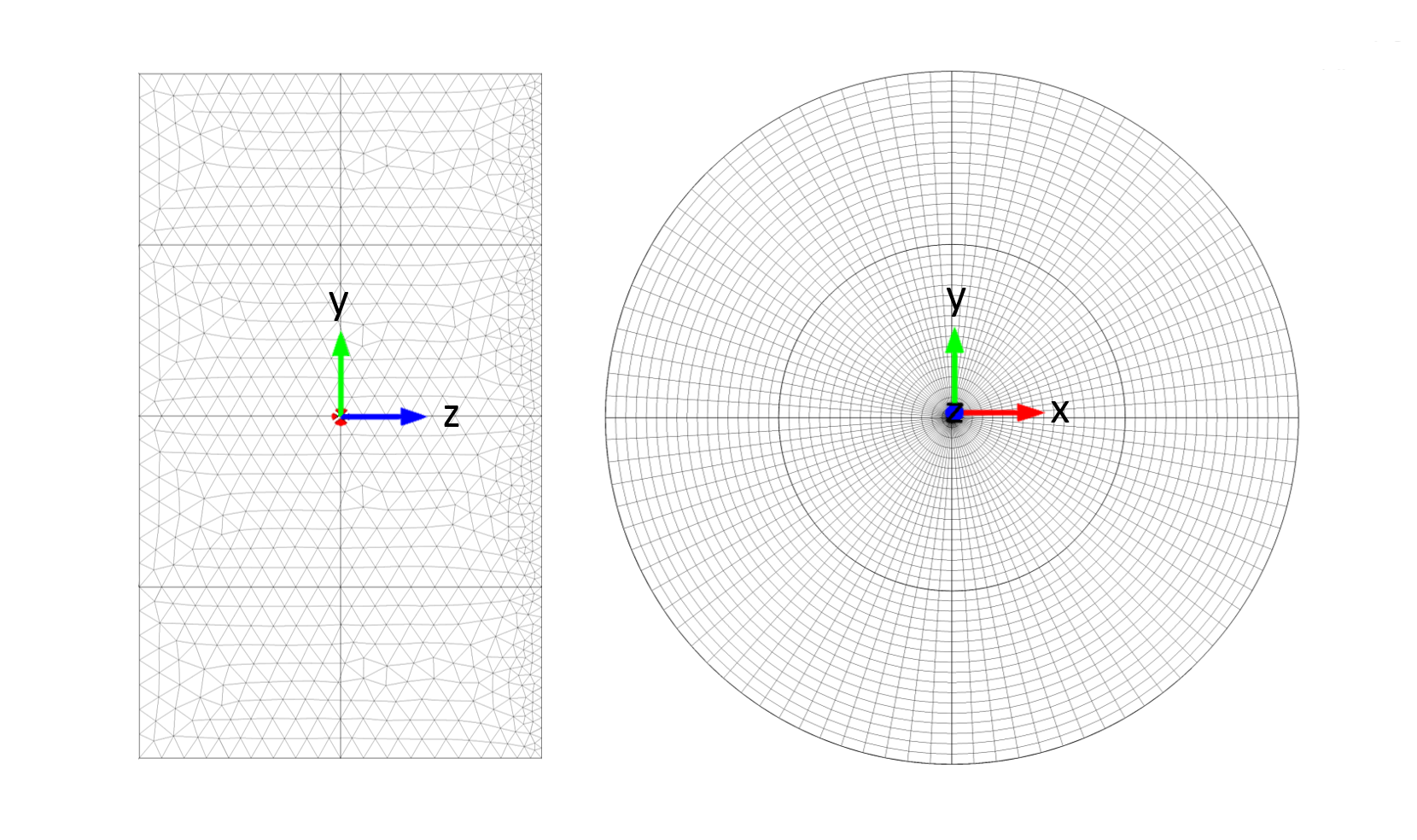}}
	\caption{ A radial cross-section of the cylindrical optic, showing the mesh used for the FEM. The mesh consisted of \(32000\) nodes, and is finest on the heated top surface of the test mass.}
	\label{figure1}
\end{figure}

We assume heating of the top face by 1 W of power absorbed with a Gaussian-distributed flux:

\begin{equation*}
Q(x,y) = \frac{2}{\pi w^2} \exp\big[-2 \big((x-x_0)^2 +(y - y_0)^2\big)/w^2 \big]
\end{equation*}

where  the beam radius \(w = 53\) mm, and radiative cooling of all surfaces of the optic to surroundings at 293 K.
A thermal FEM \cite{COMSOL} is used to calculate the temperature distribution, \(T(\vec{r})\) , resulting from the heating. The displacement amplitude for each basis function, \(a_n\) , and the total displacement, \(u_z^T(\vec{r},t)\) , are then calculated using Eq. \ref{equation4} and Eq. \ref{equation8}.

\section{Choice of orthonormal basis functions}

Choosing a set of orthonormal functions  \(\chi_n(\vec{r})\) that can describe the surface distortion without requiring a large number of functions, which would necessarily include high spatial frequencies, is crucial as it reduces both the number of auxiliary tractions that must be evaluated and the requirement for using a fine mesh in the FEM. Thus, we describe the choice of basis functions for on-axis and off-axis heating of the optic.

\subsection{Orthonormal basis for on-axis heating \( (x_0 = 0, y_0 = 0)\) }

Zernike polynomials (see Appendix A) are often used to describe cylindrically symmetric optical aberrations, as they are orthogonal over a circular disc and can be normalized. However, as shown in Figure \ref{figure2}, these polynomials are not well suited to describing the distortion.

\begin{figure}[htbp]
	\centerline{\includegraphics[width=.9\columnwidth]{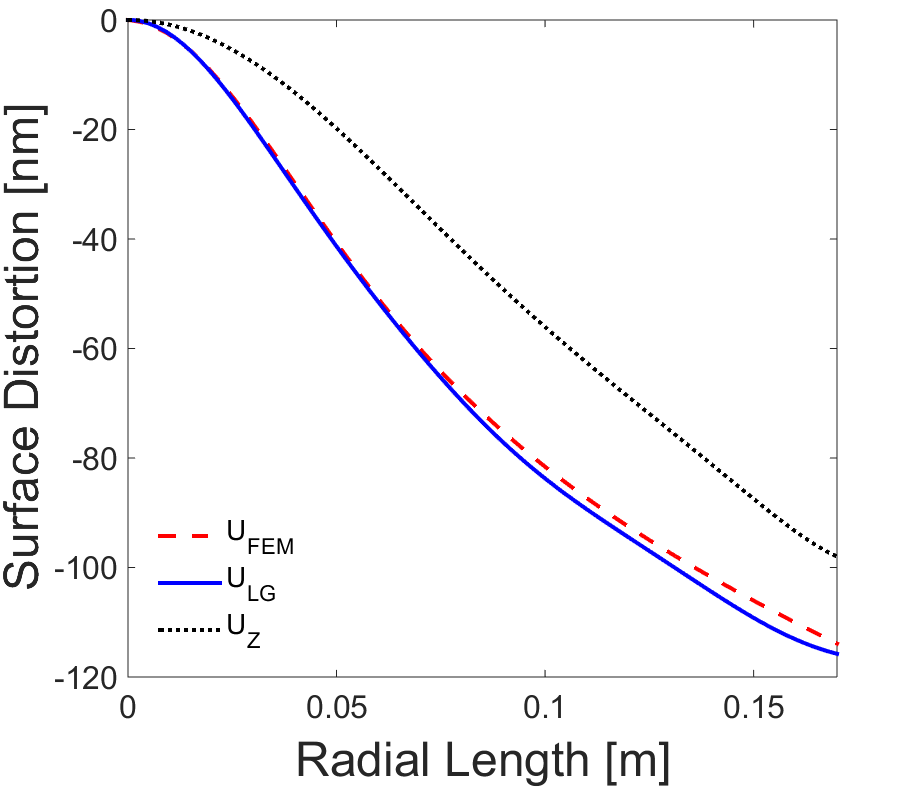}}
	\caption{Comparison of the surface distortion calculated using the elastodynamic FEM, \(u_{\textsc{\tiny FEM}}\) , the sum of the first six Zernike components  \(u_{\textsc{\tiny Z}}\), and the sum of the first six orthonormalized LG components  \(u_{\textsc{\tiny LG}}\)}
	\label{figure2}
\end{figure}

On-axis surface distortion due to the heating can also be described using Laguerre-Gauss (LG) functions:  
\begin{equation*}
LG_p(r) = L_p \Big( \frac{2 r^2}{r_0^2}\Big) \exp \Big[\frac{-r^2}{r_0^2}\Big]
\end{equation*}

where \(L_p\)  are Laguerre polynomials of order \(p:\{0,1,2...\}\)   (see Appendix A), \(r\)}  is the radial coordinate and \(r_0\) is a free parameter. These functions are orthogonal only over the infinite plane however.

Symmetric orthogonalization \cite{schweinler1970orthogonalization} is therefore used, as outlined in Appendix B, to construct linear combinations, \(\chi_n\) , of LG functions that are orthonormal over the end face for a given \(r_0\) . In this type of orthogonalization, the difference between the new and original functions is minimized in the least-squares sense \cite{schweinler1970orthogonalization}.

The optimum value of \(r_0\)   was chosen as described in Appendix C, giving \(r_0 = 1.5w\) . The six lowest-order orthonomalized-LG functions are defined in Appendix D. A comparison of \(u_{\textsc{\tiny FEM}}\)  and the sum of these components in Fig. \ref{figure2} shows that the LG basis is much superior to the Zernike basis.

\subsection{Orthonormal basis for off-axis heating}
The distortion due to off-axis heating can be described using the sets of functions listed below:

(a) Hermite-Gauss (HG) functions:
\begin{equation*}
HG_{mn}(x,y) = H_m \Big( \frac{\sqrt{2} x}{r_{0x}}\Big) \exp \Big[\frac{-x^2}{r_{0x}^2}\Big] H_n \Big( \frac{\sqrt{2} y}{r_{0y}}\Big) \exp \Big[\frac{-y^2}{r_{0y}^2}\Big]
\end{equation*}

where \(H_i\) are the (“physicists”) Hermite polynomials of order \(i:\{0,1,2,...\}\)   (see Appendix A). These functions are orthogonal over the interval \( x,y:(-\infty,\infty) \)  . We choose \(r_{0x} = r_{0y} \equiv r_{0}\)  as the heat flux has a circular cross section and we shall use \(x_0,y_0 << R \) , and thus

\begin{equation*}
HG_{mn}(x,y) = H_m \Big( \frac{\sqrt{2} x}{r_{0}}\Big) H_n \Big( \frac{\sqrt{2} y}{r_{0}}\Big) \exp \Big(\frac{-(x^2 + y^2)}{r_{0}^2}\Big) 
\end{equation*}

(b) Generalized LG functions:

\begin{equation*}
LG^l_p(r) = L_p \Big(\frac{2 r^2}{r_0^2}\Big) \exp \Big[\frac{-r^2}{r_0^2}\Big] 
\begin{cases}
1  \\
\sin{l \phi} \\
\cos{l \phi} 
\end{cases}
\end{equation*}

where \(\phi\) is the azimuthal angle, and \(l:\{1,2,3,...\}\) for \(p>0\)  . We restricted the azimuthal dependence to \(l = 1\)   due to the symmetry of the expected distortion.

Orthonormalized HG and generalized-LG functions were constructed, and an optimized value of \(r_0 = 1.4 w\)  was selected as discussed above.

HG functions up to \( m+n = 15 \)  (136 functions in total) were initially used to describe the distortion due to a heating beam that was displaced from the center of the optic according to \( (x_0,y_0)= \)(0,10 mm), (10 mm, 0)  and (8.7 mm, 5 mm). 

\begin{figure}[htbp]
	\centerline{\includegraphics[width=.9\columnwidth]{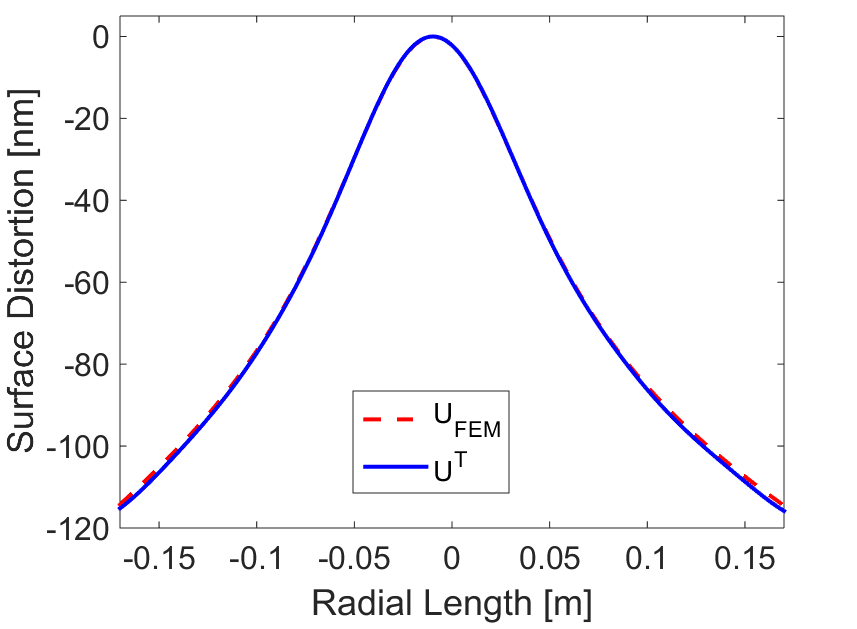}}
	\caption{Comparison of \(u_{\textsc{\tiny FEM}}\)  and the sum of the 17 dominant orthonormalized-HG components \(u^T\) for a heat flux offset of 10 mm  }
	\label{figure4}
\end{figure}

In each case, the distortion was dominated by the same 17 components, the functions for which are plotted in Appendix E. A comparison of  \(u_{\textsc{\tiny FEM}}\) and the sum of the dominant 19 components is shown in Fig. \ref{figure4} .

Orthonormalized generalized-LG functions up to \( p = 5 \)  (16 functions in total) were also generated and used to describe the distortion due to a heat flux displaced from the center of the optic by 10 mm, but they yielded slightly poorer agreement with \(u_{\textsc{\tiny FEM}}\)  . In addition, since the lower order orthonormalized-HG functions appear similar to the TEM$_{01}$ and TEM$_{10}$ eigenmodes observed in optical cavities, we chose to use that basis.

\section{Surface distortion calculated using reciprocity}

We now show how to use the orthonormal bases described above with reciprocity to determine the surface distortion. In each case, the equilibrium \( \varepsilon^A_{ij}(\vec{r})\)  values were calculated for the basis tractions and then combined with the temperature distribution \(T(\vec{r})\)  from the thermal FEM to yield the amplitudes \(a_n\).

\subsection{On-axis heating: Zernike basis}
While Zernike polynomials are not appropriate for describing the surface distortion in the example presented here, they can be used for a reciprocity-based calculation. Table I shows a comparison of the reciprocity Zernike amplitudes with those calculated by decomposing the distortion predicted by the thermoelastostatic FEM.

\begin{table}
	\begin{tabular}{ |c | c | c| }
		\hline
		Zernike polynomial & a\(_\text{n}\)(nm) & a\(_\text{n,FEM}\)(nm) \\ \hline
		Z02 & 42.6 & 42.9 \\ \hline
		Z04 & -15.2 & -15.0 \\ \hline
		Z06 & 6.3 	& 5.9  \\ \hline
		Z08 & -2.8  & -3.6 \\ \hline
	\end{tabular}
	\label{table1}
	\caption{Zernike amplitudes calculated using reciprocity, \(a_n\) , and thermoelastostatic FEM, \(a_{n,FEM}\) , for the axisymmetric Gaussian heat flux.}
\end{table}

\subsection{On-axis heating: orthonormalized-LG basis}
\begin{figure}[htbp]
	\centerline{\includegraphics[width=.9\columnwidth]{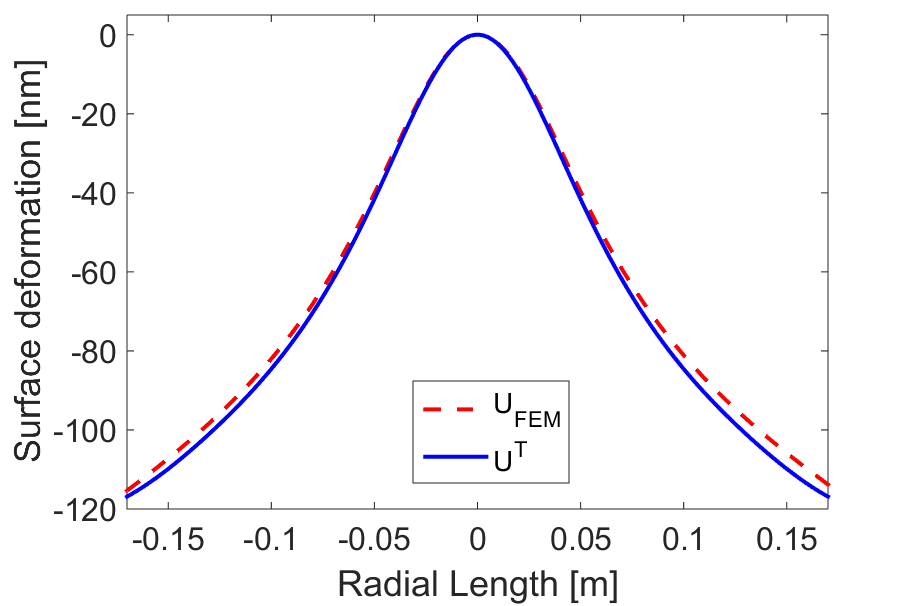}}
	\centerline{\includegraphics[width=.9\columnwidth]{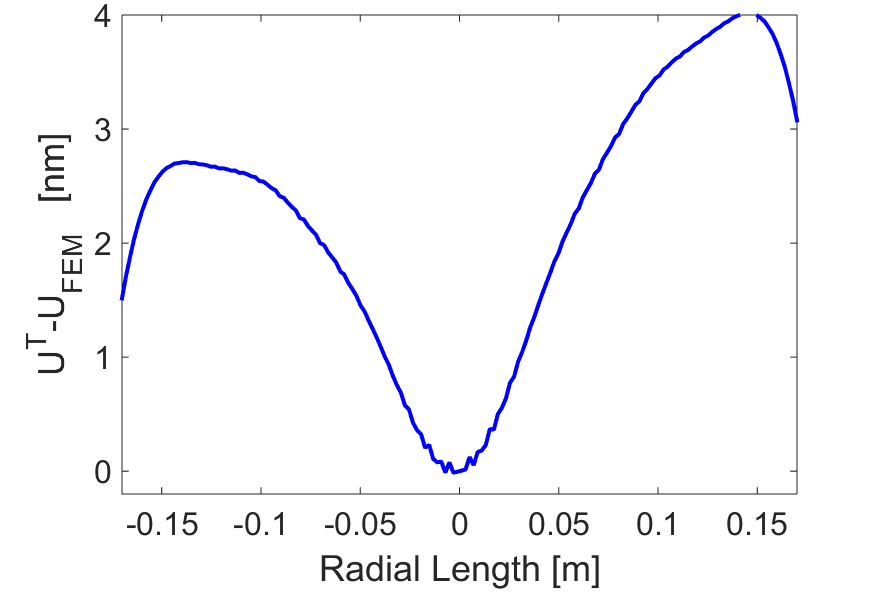}}
	\caption{ a) A plot of \(u_{\textsc{\tiny FEM}}\)  and  \(u^T\) calculated for the on-axis heating using the 6 lowest-order orthonormalized-LG functions. b) A plot of \(u_{\textsc{\tiny FEM}} - u^T\) . }
	\label{figure5}
\end{figure}

The \(u_{\textsc{\tiny FEM}}\) and  \(u^T = \sum\limits_{n=0}^5 a_n \chi_n \), and the difference between the two curves are plotted in Fig. \ref{figure5}. Since we are not interested in the average displacement of the optics, we have set \(u^T =u_{\textsc{\tiny FEM}} \) at $r=0$.   The asymmetry of the difference is due to non-ideal cylindrical symmetry in the FEM meshing.

\subsection{Off-axis heating: orthonormalized-HG basis}

\begin{figure}[htbp]
	\centerline{\includegraphics[width=.9\columnwidth]{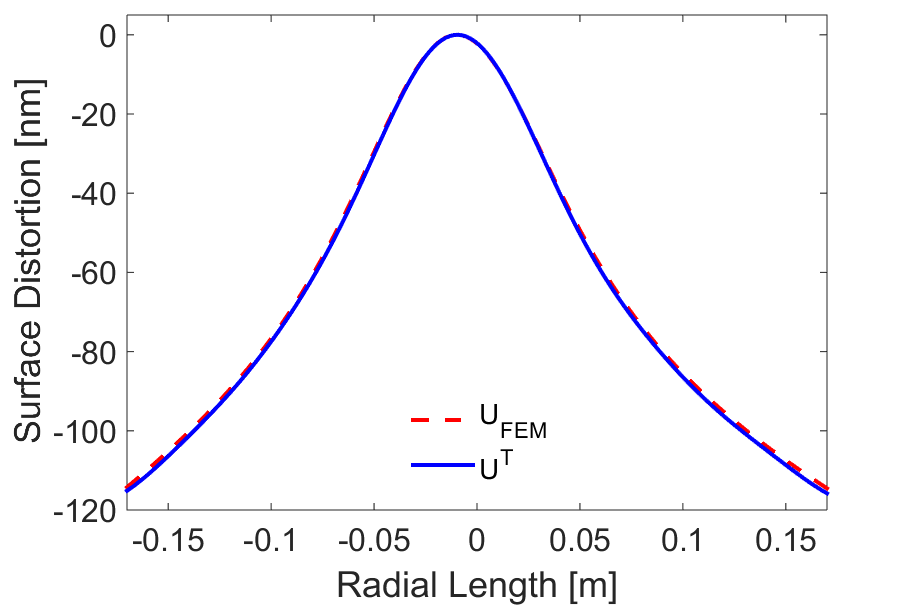}}
	\centerline{\includegraphics[width=.9\columnwidth]{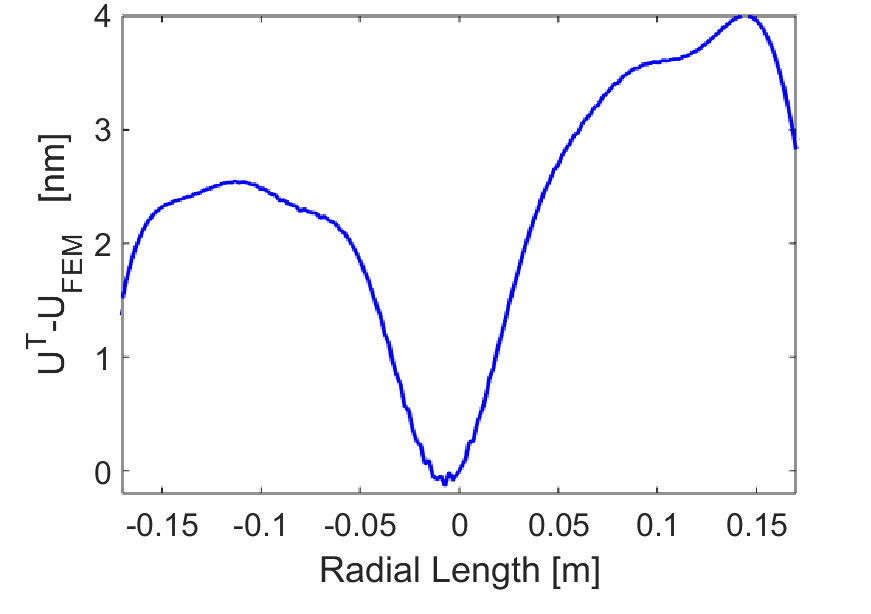}}
	\caption{a) A plot of \(u_{\textsc{\tiny FEM}}\) and \(u^T\) calculated for the off-axis heating using the 17 dominant orthonormalized-HG functions. b) A plot of \(u_{\textsc{\tiny FEM}} - u^T\).}
	\label{figure6}
\end{figure}

The \(u_{\textsc{\tiny FEM}}\) and \(u^T = \sum\limits_{n = 1}^{19} a_n \chi_n \) and the difference between the two curves are plotted in Fig. \ref{figure6}

Figures \ref{figure5} and \ref{figure6} show that even though $<20$ auxiliary tractions were used to characterize the optic and the FEM was restricted to only 30,000 nodes,

\begin{itemize}
	\item Elastodynamic reciprocity predicts $u^T$ within $ <1.5 \% $  of \(u_{\textsc{\tiny FEM}}\) over the majority of the incident laser beam
	\item Displacing the beam by $20\%$ of its radius does not degrade the agreement.
\end{itemize}
Additionally, increasing the number of auxiliary tractions further improves the agreement, particularly at large radius.

\section{Comparison of Computational Times}

We compare here the times required to calculate the surface distortion using our hybrid FEM-reciprocity approach and using a conventional thermo-elastic FEM analysis. The times are specific to the example of partial absorption of a Gaussian-intensity-profile light beam by the surface of an isotropic cylindrical optic.

In both cases we use 32,000 nodes in the FEM calculations. We have not yet investigated how many nodes or auxiliary tractions are required to achieve a particular accuracy for each approach, or how this might affect the computational times.

As discussed earlier, our hybrid FEM-reciprocity approach consists of two parts, the first of which is done only once for an optic:

\begin{enumerate} 
	\item 
	\begin{enumerate} 
		\item Calculate the elastic response of the optic to each of the orthonormal tractions, and store these arrays in memory. Here, this consisted of a 32,000 long 6-element array in which the 3D coordinates and strains at each node were recorded for each traction. This part required approximately 1 hour per traction.
		\item Upload 20 responses into memory in preparation for part 2 required 20 minutes. 
	\end{enumerate}
	\item At each epoch of interest
	\begin{enumerate} 
	\item Calculate the thermal induced stress at each node using FEM: 90 seconds
	\item Evaluate the volume integral for each traction component using Eq. (6): 3 seconds per traction. Thus for a serial calculation with 20 tractions, this step required 60 seconds.
	\end{enumerate}
\end{enumerate}

A conventional thermo-elastic FEM calculation for this simple problem required about 13 minutes.

Thus, once the response of the optic has been determined and uploaded, the hybrid FEM-reciprocity calculation is at 5.2 times faster is using a serial calculation, and 8.7 times faster if using a parallelized calculation of the distortion using reciprocity.

\section{Conclusion}
We have shown how Betti-Maxwell reciprocity can be used in combination with thermal finite-element modeling to calculate the thermoelastic distortion of a linear elastic system. As an example, we described in detail its application to calculating the distortion of the end face of an isotropic cylindrical glass optic heated by an off-axis Gaussian laser beam. Despite using less than 20 auxiliary eigenfunction tractions to characterize the optic, the distortion calculated using reciprocity agrees to $<1.5 $ \% with that calculated using a full thermoelastic FEM over the majority of the incident beam.

The computational time required for the reciprocity approach was a factor of 5-8 less than that for the full FEM once the optic had been characterized. The advantage of this approach will thus be most evident in cases where the elastic distortion must be calculated frequently, such as in feed-forward control of systems with long thermal time constants for example. Parallelization of the reciprocity calculation would also allow further improvements to the accuracy by employing additional tractions but with negligible additional computational cost.

Our reciprocity approach can be applied to systems with arbitrarily distributed heat fluxes and asymmetric anisotropic elastic bodies. Furthermore, while our example assumed a free optic, other boundary conditions could easily be incorporated into the analysis with an appropriate set of auxiliary eigenfunctions.

\section{Acknowledgements}
This research was supported by the Australian Research Council (ARC). Yuri Levin is supported by an ARC Future Fellowship

\appendix
\section{Polynomials used in this paper}
\begin{table}[!htbp]
	\begin{tabular}{ |c | c |  }
		\hline
		Zernike Polynomial 	& 	orthogonal form with \(\rho = r/R)\) \\ \hline
		Z02 				&   \(\sqrt{\frac{3}{\pi}} (2 \rho^2 -1)\) \\ \hline
		Z04 				& 	\(\sqrt{\frac{5}{\pi}} (6 \rho^4 - 6 \rho^2 + 1 ) \) \\ \hline
		Z06					&   \(\sqrt{\frac{7}{\pi}} (20 \rho^6 - 30 \rho^4 + 12 \rho^2 -1) \) \\ \hline
		Z08					&  	\(\sqrt{\frac{9}{\pi} } (70\rho^8 - 140 \rho^6 + 90 \rho^4 - 20 \rho^2 +1) \) \\ \hline
	\end{tabular}
	\label{tableA1}
	\caption{Zernike amplitudes calculated using reciprocity,\(a_n\) , and thermoelastostatic FEM, \(a_{n,FEM}\) , for the axisymmetric Gaussian heat flux.}
\end{table}

\begin{table}[!htbp]
\begin{tabular}{ |c | c | c| }
	\hline
	$n$	& 	\(L_n(x)\)								  &	\(H_n(x)\) 				\\  \hline	
	0	&	1										  &	   1					\\ \hline
	1	&	\( - X + 1\)							  & 	\(2 x\)				\\ \hline
	2	&	\((x^2 - 4x +2)/2\)						  &	\(4 x^2 -2\) 			\\ \hline
	3	&	\((-x^3 + 9x^2 -18x +6)/6\)				  &	\(8 x^3 - 12x\)			\\ \hline
	4	& 	\((x^4 - 16x^3 + 72 x^2 - 96 x +24)/24 \) &	\(16x^4 - 48 x^2 + 12\) \\ \hline
\end{tabular}
	\caption{Laguerre and Hermite polynomials used.}
\end{table}

\section{Summary of symmetric orthogonalization}
The linearly independent LG and HG functions, denoted here by \(f_k(\vec{r})\)  , were orthonormalized over the end face of the mirror using the following process \cite{schweinler1970orthogonalization}:

\begin{enumerate}
\item Calculate the matrix of inner products of the functions: \(M_{kl} = \iint \limits_{end \ face} f_k(\vec{r})f_l(\vec{r})dS\)  where the integration was evaluated for the mesh used to export the data from the FEM. In this work, the data was exported on a 1mm-pitch mesh and cropped to fit within the circular endface of the optic.

\item Determine the eigenvalues \(p_k\)  and eigenvectors \(\vec{u}_\lambda\)  of the inner product matrix such that \( \sum_l M_{kl}  u_{l \lambda}  = p_\lambda u_{k \lambda} \)  

\item	The orthonormalized functions \(\chi(\vec{r})\)  are then given by \(\chi_n = \frac{1}{\sqrt{p_n}} \sum\limits_{k} u_{kn} f_k\)
\end{enumerate}

\section{Optimization of \(r_0\)}
The optimum $r_0$  was chosen to minimize the mean squared difference, weighted by the amplitude of the incident laser beam, between \(u_{\textsc{\tiny FEM}}\) and the sum of the selected orthonormalised components using:

\begin{equation*}
\frac{\iint \limits_{end \ face} \Big( u_{FEM} - \sum\limits^5_{n = 0} a_n \chi_n \Big)^2 \exp \Big[\frac{(x-x_0)^2 + (y-y_0)^2}{w^2}\Big]dS}{\iint \limits_{end \ face}  \exp \Big[\frac{(x-x_0)^2 + (y-y_0)^2}{w^2}\Big]dS}
\end{equation*}

where

\begin{equation*}
a_n = \iint\limits_{end \ face} u_{\textsc{\tiny FEM}} \chi_n dS	
\end{equation*}
	
and a new orthonormal set of functions \(\chi_n\) was generated for each value of \(r_0\)  . The integrations were evaluated using a square array of pitch 1 mm within the end face.

\begin{figure}[htbp]
	\centerline{\includegraphics[width=.9\columnwidth]{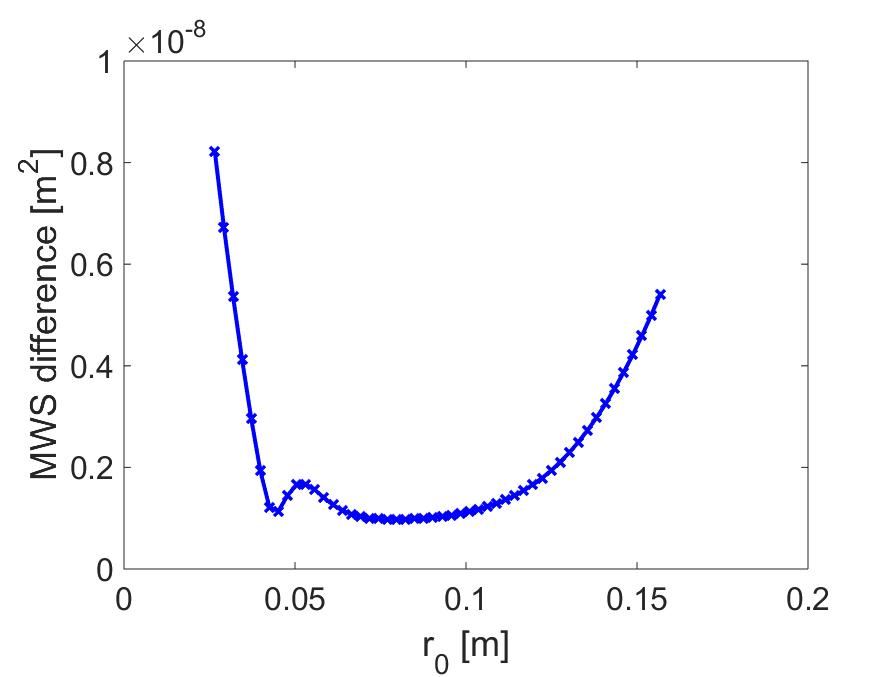}}
	\caption{Plot of the mean-weighted-squared difference between \(u_{\textsc{\tiny FEM}}\) and the sum of the first six orthonormalized-LG components as a function of $r_0$ .}
	\label{figureC1}
\end{figure}

The variation of this mean-weighted-squared difference with \(r_0\) for the axisymmetric heating   \((y_0 = 0) \)\ is plotted in Figure \ref{figureC1}.

\begin{figure}[htbp]
	\centerline{\includegraphics[width=.9\columnwidth]{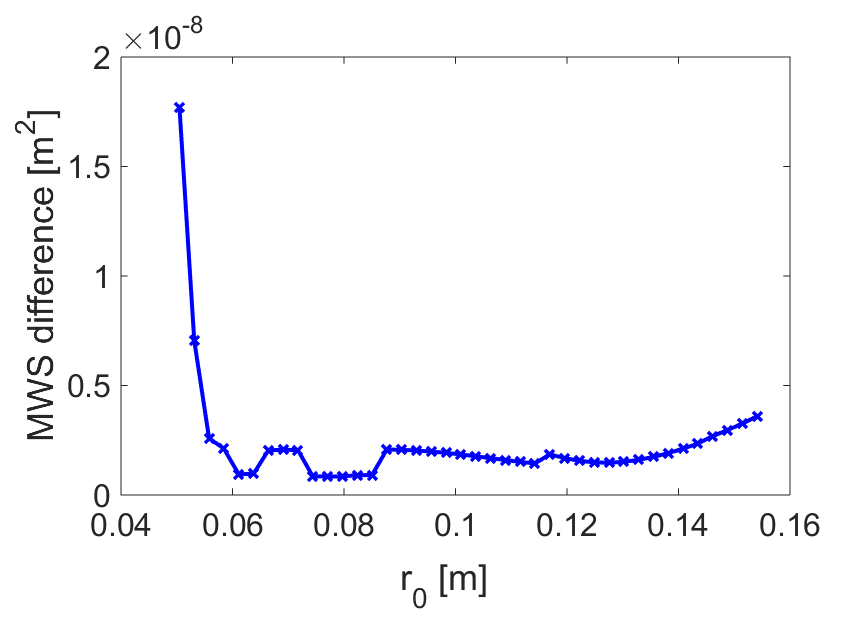}}
	\caption{Plot of the mean-weighted-squared difference between \(u_{\textsc{\tiny FEM}}\) and the sum of the first six orthonormalized-HG components as a function of $r_0$ .}
	\label{figureC2}
\end{figure}

The variation of this mean-weighted-squared difference with $r_0$ for orthonormalized-HG functions and off-axis heating \(y_0 = \) 10 mm   is plotted in Fig. \ref{figureC2}.

\section{Orthonormalized-LG functions used in this paper}

\begin{multline*}
\chi_n = c_{0n}LG_0 + c_{1n}LG_1 + c_{2n}LG_2 \\
+ c_{3n}LG_3 + c_{4n}LG_4 + c_{5n}LG_5
\end{multline*}

\begin{table}[!htbp]
	\begin{tabular}{|c|c|c|c|c|c|c|}
		\hline
		$n$ & $c_{0n}$ & $c_{1n}$ & $c_{2n}$ & $c_{3n}$ & $c_{4n}$ & $c_{5n}$ \\ \hline
		0 & -0.97 & -0.25 & -0.06 & -0.01 & -0.001 & -0.0001 \\ \hline
		1 & 0.24  & -0.86 & -0.43 & -0.13 & -0.028 & -0.003  \\ \hline
		2 & 0.062 & -0.42 & 0.69  & 0.55  & 0.20   & 0.039   \\ \hline
		3 & 0.016 & -0.15 & 0.52  & -0.53 & -0.62  & -0.24   \\ \hline
		4 & 0.005 & -0.055 & 0.26 & -0.58 & 0.40  & 0.79 \\ \hline
		5 & -0.005 & 0.07 & -0.4 & 1.23 & -2.24 & 2.17 \\ \hline
		 
	\end{tabular}
\end{table}

\section{The 17 dominant orthonormalized-HG functions}

\begin{figure}[h]
	\begin{center} 
		\includegraphics[width=50mm]{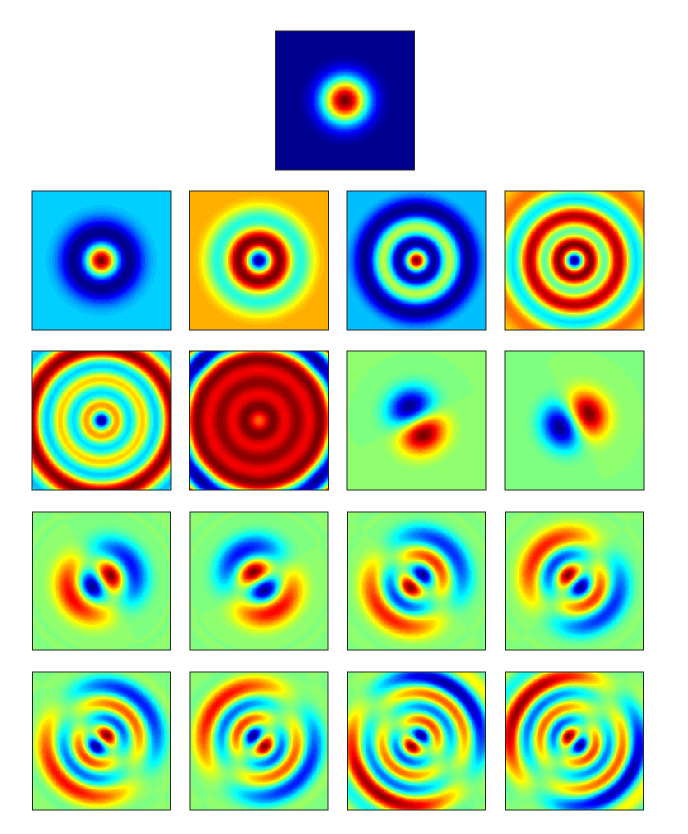}
	\end{center}
	\caption{The 17 dominant orthonormalized-HG functions}
	\label{figureC3}
\end{figure}

The Hermite Gauss functions up to order \( n+m=15 \) are orthogonalized using symmetric orthogonaliazation.  Of these 136 modes, the 17 modes that make the largest contribution to describing the deformed surface were selected.  These modes are shown in Fig. \ref{figureC3}

\begin{thebibliography}{99}
	

\bibitem{2015advancedligo}
Aasi, J., \emph{et al}. (LIGO Scientific Collaboration) Class. Quant. Grav. \textbf{32}, 074001 (2015).


\bibitem{virgo2009advanced}
Virgo Collaboration \emph{et al.} , ``Advanced virgo baseline design,''
(2009) VIR-027A-09 The Virgo Collaboration.

\bibitem{somiya2012detector}
K. Somiya, Class. Quant. Grav. \textbf{29},
124007 (2012).

\bibitem{lawrence2003active}
R. C. Lawrence, Ph.D. thesis, Massachusetts Institute of
Technology (2003).

\bibitem{boley2012theory}
B. A. Boley and J. H. Weiner, \emph{Theory of thermal stresses}, Courier Dover
Publications, (2012).

\bibitem{hello1990analytical}
P. Hello and J.-Y. Vinet, J. Phys. \textbf{51}, 1267(1990).

\bibitem{achenbach2006reciprocity}
J. D. Achenbach, ``Reciprocity and related topics in elastodynamics,''
Appl. Mech. Rev. \textbf{59}, 13 (2006).

\bibitem{achenbach2005thermoelasticity}
J. D. Achenbach, J. Therm. Stress. \textbf{28}, 713 (2005).

\bibitem{achenbach2007application}
J. D. Achenbach, J. Therm. Stress. \textbf{30}, 841 (2007).

\bibitem{levin2012creep}
Y. Levin, Phys. Rev. D \textbf{86}, 122004 (2012).

\bibitem{COMSOL}
``Comsol multiphysics''.

\bibitem{schweinler1970orthogonalization}
H. Schweinler and E. Wigner, J. Math. Phys. \textbf{11}, 1693(1970).

\bibitem{vinet2004virgo}
J. Vinet \emph{et al.}, ``The Virgo physics book'', available online at:
http://wwwcascina.virgo.infn.it/vpb .

\end{thebibliography}
\end{document}